\documentclass[11pt]{article}

\usepackage{amsthm}
\usepackage[hyperindex,breaklinks]{hyperref}
\RequirePackage[authoryear]{natbib}
\usepackage{xcolor}

\hypersetup{colorlinks=true,
  citecolor=[HTML]{215E21},
  urlcolor=[HTML]{0050D0},
  linkcolor=[HTML]{E01939}}

\usepackage{breakcites}

\usepackage[font=small,labelfont=bf]{caption}
\bibpunct[ ]{(}{)}{;}{a}{}{;}
\usepackage[utf8]{inputenc} 

\usepackage[english]{babel}
\usepackage[autostyle]{csquotes}

\usepackage{algorithm}
\usepackage{algorithmic}

\usepackage{amsmath}
\usepackage{amsfonts}
\usepackage{amssymb}
\usepackage{mathtools}
\mathtoolsset{showonlyrefs}

\usepackage{authblk}
\usepackage{footmisc}

\newcommand{\dotp}[2]{\ensuremath{\left< #1 , #2 \right>}}

\newcommand{\eqdef}{\coloneqq}

\newcommand{\newsubsubsection}[1]{\noindent\textbf{#1.}~}

\DeclareMathOperator*{\argminOp}{argmin}

\DeclareMathOperator{\diagOperator}{diag}

\DeclareMathOperator{\traceOperator}{tr}

\DeclareMathOperator{\ot}{OT}

\newcommand{\argmin}[1]{\argminOp_{#1}}
\newcommand{\trace}[1]{\traceOperator\left( #1 \right)}
\newcommand{\diag}[1]{\diagOperator\left( #1 \right)}

\renewcommand{\vector}[1]{\ensuremath{\lowercase{\boldsymbol{#1}}}}
\newcommand{\row}[2]{\vector{#1}_{#2:}}
\newcommand{\col}[2]{\vector{#1}_{#2}}
\newcommand{\ones}{\vector{1}}
\newcommand{\dataseturl}{\url{http://www.ntt-at.com/product/speech2002/}}

\begin{document}

\title{Blind Source Separation with Optimal Transport 
Non-negative Matrix Factorization}

\newcommand*\samethanks[1][\value{footnote}]{\footnotemark[#1]}
\author[1]{Antoine Rolet\thanks{Work performed during an internship at NTT Communication Science Laboratories}}
\author[1]{Vivien Seguy\samethanks}
\author[2]{Mathieu Blondel}
\author[2]{Hiroshi Sawada}

\affil[1]{Graduate School of Informatics\\ Kyoto University}
\affil[2]{NTT Communication Science Laboratories}
\date{}

\maketitle

\begin{abstract} 
Optimal transport as a loss for machine learning optimization problems has
recently gained a lot of attention. Building upon recent advances in
computational optimal transport, we develop an optimal transport non-negative
matrix factorization (NMF) algorithm for supervised speech blind source
separation (BSS).  Optimal transport allows us to design and leverage a cost
between short-time Fourier transform (STFT) spectrogram frequencies, which takes
into account how humans perceive sound.  We give empirical evidence that using
our proposed optimal transport NMF leads to perceptually better results than
Euclidean NMF, for both isolated voice reconstruction and BSS tasks.  Finally,
we demonstrate how to use optimal transport for cross domain sound processing
tasks, where frequencies represented in the input spectrograms may be different
from one spectrogram to another.  
\end{abstract}

\section{Introduction}

Blind source separation (BSS) is the task of separating a mixed signal into 
different components, usually referred to as sources. In the context of sound processing, it 
can be used to separate speakers whose voices have been recorded simultaneously.
A common way to address this task is to decompose the signal spectrogram 
by non-negative matrix  factorization \citep[NMF,][]{lee2001algorithms}, as
proposed for example by \citet{schmidt2006single} as well as \citet{sun2013universal}. Denoting 
$\tilde{x}_{j, i}$ the (complex) short-time Fourier transform (STFT) coefficient of the input signal at 
frequency bin $j$ and time frame $i$, and $X$ its magnitude spectrogram defined as $x_{j, i} = |\tilde{x}_{j, i}|$, 
the BSS problem can be tackled by solving the NMF problem
\begin{equation}
\label{eq:nmf_bss}
\min_{D^{(1)}\dots D^{(N)},\, W^{(1)}\dots W^{(N)}} 
\sum_{i=1}^t\ell\left(\col{X}{i}, \sum_{k=1}^ND^{(k)}\col{W}{i}^{(k)}\right)
\end{equation}
where $N$ is the number of sources, $t$ is the number of time windows,
$\col{X}{i}$ is the $i$\textsuperscript{th} column of $X$ and $\ell$ is a loss function.
Each dictionary matrix $D^{(k)}$ and 
weight matrix $W^{(k)}$ are related to a single source. In a supervised 
setting, each source has training data and all the $D^{(k)}$s are learned in 
advance during a training phase. At test time, given a new signal, 
separated spectrograms are recovered from the $D^{(k)}$s and 
$W^{(k)}$s and corresponding signals can be reconstructed with suitable 
post-processing. Several loss functions $\ell$ have been considered in the literature, such as the squared Euclidean distance~\citep{lee2001algorithms,schmidt2006single}, 
the Kullback-Leibler divergence~\citep{lee2001algorithms,sun2013universal} or the Itakura-Saito 
divergence~\citep{fevotte2009nonnegative,sawada2013multichannel}.

\begin{figure*}[h!]
\includegraphics[width=\textwidth]{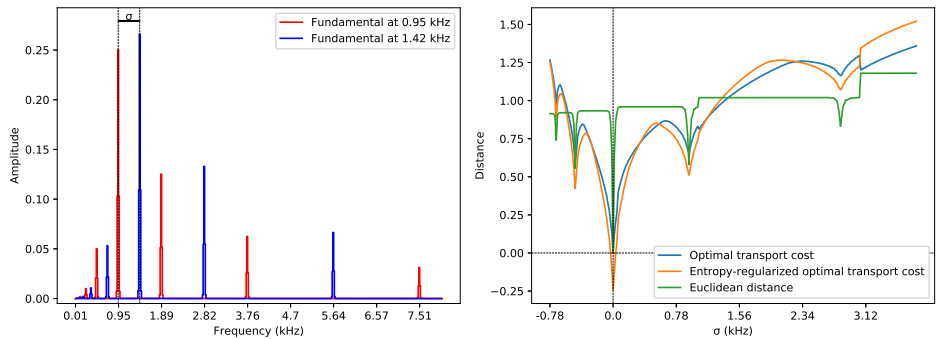}\\
  \caption{\textbf{Comparison of Euclidean distance and (regularized)
  optimal transport losses.} Synthetic musical notes are generated by
  putting weight on a fundamental, and exponentially decreasing weights on its
  harmonics and sub-harmonics, and finally convoluting with a Gaussian. Left:
  examples of the spectrograms of two such notes. Right: (regularized)
  optimal transport loss and Euclidean distance from the note of
  fundamental $0.95$kHz (red line on the left plot) to the note of fundamental
  $0.95$kHz$+\sigma$, as functions of $\sigma$. The Euclidean distance varies
  sharply whereas the optimal transport loss captures more smoothly the
  change in the fundamental. The variations of the optimal transport
  loss and its regularized version are similar, although the regularized one can become negative.}
  \label{fig:intro}
\end{figure*}

In the present article, we propose to use \textit{optimal transport} as a loss between spectrograms to
perform supervised speech BSS with NMF. Optimal transport is defined 
as the minimum cost of moving the mass from one histogram to another.
By taking into account a transportation cost between frequencies, this provides a powerful
metric to compare STFT spectrograms. One of the main advantage of using 
optimal transport as a loss is that it can quantify the amplitude of a frequency shift
noise, coming for example from quantization or the tuning of a musical instrument.
Other metrics such as the Euclidean distance or Kullback-Leibler divergence,
which compare spectrograms element-wise, are almost blind to this type of noise (see Figure \ref{fig:intro}). 
Another advantage over element-wise metrics is that optimal transport
enables the use of different 
quantizations, i.e. frequency supports, at training and test times. Indeed, the frequencies represented on a 
spectrogram depend on the sampling rate of the signal and the time-windows used 
for its computation, both of which can change between training and test times. With optimal 
transport, we do not need to re-quantize the training and testing data so as
they share the same frequency support: optimal transport is well-defined between spectrograms 
with distinct supports as long as we can define a transportation cost between 
frequencies. Finally, the optimal transport framework enables us to generalize the Wiener filter, a common post-processing 
for source separation, by using optimal transport
plans, so that it can be applied to data quantized on different frequencies.

NMF with an optimal transport loss was first proposed by~\cite{sandler09}. They
solved this problem by using a bi-convex formulation and relied on 
an approximation of optimal transport
based on wavelets~\citep{Shirdhonkar08}. Recently, \cite{rolet2016} proposed fast algorithms
to compute NMF with an entropy-regularized optimal transport loss, which are
more flexible in the sense that they do not require any assumption on the frequency quantization
or on the cost function used.

Using optimal transport as a loss between spectrograms was also proposed by 
\cite{flamary2016optimal} under the name \enquote{optimal spectral transportation}. They
developed a novel method for unsupervised music transcription which achieves state-of-the-art performance.
Their method relies on a cost matrix designed specifically for musical
instruments, allowing them to use Diracs as dictionary columns.
That is, they \textit{fix} each dictionary column to a vector with a single
non-zero entry and learn only the corresponding coefficients.
This trivial structure of the dictionary results in
efficient coefficient computation.
However, this approach cannot be applied as is to speech 
separation since it relies on the assumption that a musical note can be represented
as its fundamental. It also requires designing 
the cost of moving the fundamental to its 
harmonics and neighboring frequencies. Because human voices are intrinsically more
complex, it is therefore necessary to learn \textit{both} the dictionary and the
coefficients, i.e., solve full NMF problems.

\subsection*{Our contributions}
In this paper, we extend the optimal transport NMF of \cite{rolet2016} to the case where
the columns of the input matrix $X$ are not normalized in order to propose an algorithm which
is suitable for spectrogram data. Normalizing all time frames so that they have the
same total weight is not desirable in sound processing tasks because it would amplify noise. 
We define a cost between frequencies so that the optimal transport objective between spectrograms
provides a relevant metric between them. We apply our NMF framework to single voice reconstruction and 
blind source separation and show that an optimal transport loss provides better results over the usual squared Euclidean loss.
Finally, we show how to use our framework for cross domain BSS, where frequencies represented in the
test spectrograms may be different from the ones in the dictionary. This may happen for example when
train and test data are recorded with different equipment, or when the STFT is computed with different parameters.

\subsection*{Notations}

We denote matrices in upper-case, vectors in bold lower-case and scalars in 
lower-case. If $M$ is a matrix, $M^\top$ is its transpose, $\col{M}{i}$ is its
$i$\textsuperscript{th} column and 
$\row{M}{j}$ its $j$\textsuperscript{th} row. $\ones_n$ denotes the all-ones vector in 
$\mathbb{R}^n$; when the dimension can be deduced from context we simply write 
$\ones$. For two matrices $A$ and $B$ of the same size, we denote their
inner product $\dotp{A}{B}\coloneqq\trace{A^\top B}$. We denote $\Sigma_n$
the $(n-1)$-dimensional simplex: $\Sigma_n
\coloneqq \left\{ \vector{x}\in\mathbb{R}_+^n \colon \|\vector{x}\|_1=1
\right\}$.

\section{Background}

We start by introducing optimal transport, its entropy regularization, which we will use as the
loss $\ell$, and previous works on optimal transport NMF. For a more
comprehensive overview of
optimal transport from a computational perspective, see \cite{peyre2017computational}.

\subsection{Optimal Transport}

\newsubsubsection{Exact Optimal Transport} Let $\vector{a}\in\Sigma_m,\, \vector{b}\in \Sigma_n$. The polytope of transportation matrices 
between \vector{a} and \vector{b} is defined as
\begin{equation*}
U(\vector{a}, \vector{b}) \coloneqq \left\{ T \in \mathbb{R}_+^{m \times n} \left| 
\begin{aligned}
T \ones = \vector{a} \\
T^\top \ones = \vector{b}
\end{aligned} \right.\right\}.
\end{equation*}
Given a cost matrix $C 
\in \mathbb{R}^{m \times n}$, the minimum transportation cost
between \vector{a} and \vector{b} is defined as
\begin{equation}
\label{eq:ot_normalized}
\ot(\vector{a},\vector{b}) = \min_{T \in U(\vector{a},\vector{b})} \dotp{T}{C}.
\end{equation}
When $n=m$ and the cost matrix is the $p$-th power ($p\geqslant 1$) of a distance matrix, i.e.
$c_{i,j} = d(\vector{y}_i, \vector{y}_j)^p$ for some $(\vector{y}_i)$ in a metric space $(\Omega,\, d)$, 
then $\ot(\cdot,\cdot)^{1/p}$ is a distance on the set of vectors in 
$\mathbb{R}_+^n$ with the same $\ell$-$1$ norm \citep[Theorem~7.3]{villani2003topics}. 
We can see the vectors $\vector{y}_i$ as features, and $\vector{a}$ and $\vector{b}$ as the quantization 
weights of the data onto these features. In sound processing applications, the vectors
$\vector{y}_i$ are real numbers corresponding to the frequencies of the
spectrogram and $\vector{a}$ and $\vector{b}$ are their corresponding magnitude. By computing
the minimal transportation cost between frequencies of two spectrograms, optimal
transport exhibits variations in accordance with the frequency noise involved in the signal generative process,
which results for instance from the tuning of musical instruments or the subject's 
condition in speech processing.

\newsubsubsection{Unnormalized Optimal Transport} In this work, we wish to define optimal transport
when $\vector{a}$ and $\vector{b}$ are non-negative but not necessarily normalized. Note that the transportation 
polytope is not empty as long as $\vector{a}$ and $\vector{b}$ sum to the same value: 
$U(\vector{a},\vector{b})=\emptyset$ \textit{iif} $\|\vector{a}\|_1 \neq
\|\vector{b}\|_1$.
Hence, we define optimal transport between possibly unnormalized vectors $\vector{a}$ and $\vector{b}$ as,
\begin{equation}
\label{eq:ot}
\ot(\vector{a},\vector{b}) \coloneqq \left\{\begin{aligned}
&\min_{T \in U(\vector{a},\vector{b})} \dotp{T}{C} &&\mbox{if }\|\vector{a}\|_1 
= \|\vector{b}\|_1\mbox{ and }\vector{a},\vector{b}\geq 0;\\
&\infty &&\mbox{otherwise}.
\end{aligned}\right.
\end{equation}

Computing the optimal transport cost \eqref{eq:ot} amounts to solve a linear program
(LP) which can be done with specialized versions of the simplex algorithm with worst-case 
complexity in $\mathcal{O}(n^3\log n)$ when $n=m$~\citep{Orlin97Polynomial}. When considering 
$\ot$ as a loss between histograms supported on more than a few hundred bins, such computation
becomes quickly intractable. 
Moreover, using $\ot$ as a loss involves
differentiating $\ot$,
which is not differentiable everywhere. Hence, one would 
have to resort to subgradient methods. This would be prohibitively slow since
each iteration would require to obtain a subgradient at the current iterate,
which requires to solve the LP \eqref{eq:ot}.

\newsubsubsection{Entropy Regularized Optimal Transport} 
To remedy these limitations, \cite{cuturi13} proposed to add an entropy-regularization term
to the optimal transport objective, thus making the $\ot$ loss differentiable
everywhere and strictly convex. This \textit{entropy-regularized optimal transport}
has since been used in numerous works as a loss for diverse tasks \citep[see for example][]{gramfort2015fast,frogner2015learning,rolet2016}.

Let $\gamma > 0$, 
we define the (unnormalized) entropy-regularized OT between 
$\vector{a}\in\mathbb{R}_+^m,\, \vector{b}\in \mathbb{R}_+^n$ as
\begin{equation}
\label{eq:ot_reg}
\ot_{\gamma}(\vector{a},\vector{b}) \coloneqq \left\{\begin{aligned}
&\min_{T \in U(\vector{a},\vector{b})} &&\dotp{T}{C} - \gamma E(T) &&\mbox{ if}\quad \|\vector{a}\|_1 
= \|\vector{b}\|_1\mbox{ and }\vector{a},\vector{b}\geq 0;\\
&&&\infty &&\mbox{otherwise.}
\end{aligned}\right.
\end{equation}
where $E(T) \coloneqq \sum_{ij}T_{ij}\log{T_{ij}}$ is the entropy of the transport plan $T$.
Let us denote $\ot_{\gamma}^\star$ the convex conjugate of $\ot_{\gamma}$ with respect to
its second variable 
$$\ot_{\gamma}^\star(\vector{x},\vector{y})=\max_{\substack{\vector{z}
\geq0\\\|\vector{z}\|_1=\|\vector{x}\|_1}}\dotp{\vector{y}}{\vector{z}}
- \ot_{\gamma}(\vector{x}, \vector{z}).$$
\cite{cuturi2016smoothed} showed that its value and gradient can be
computed in closed-form:
\begin{align*}
&\ot_{\gamma}^\star(\vector{x},\vector{y}) = \gamma\left( 
E(\vector{x})+\dotp{\vector{x}}{\log K\vector{\alpha}}\right), \\
& \nabla_{\vector{y}}\ot_{\gamma}^\star(\vector{x},\vector{y}) =\vector{\alpha} \odot \left( 
K^\top\frac{\vector{x}}{K\vector{\alpha}}\right),
\end{align*}
where $K\eqdef e^{- C/\gamma}$ and $\vector{\alpha} \eqdef e^{ 
\vector{y}/\gamma}$.

\subsection{Optimal Transport NMF}

NMF can be cast as an optimization problem of the form
\begin{equation}
\label{eq:nmf}
\min_{D\in\mathbb{R}_+^{n\times k},\, W\in\mathbb{R}_+^{k\times t}} 
\sum_{i=1}^t\ell(\col{x}{i}, D\col{w}{i}) + R(W,D),
\end{equation}
where both $D$ and $W$ are optimized at train time, and $D$ is fixed at test time.
When $\ell$ is $\ot$, problem~\eqref{eq:nmf} is convex 
in $W$ and $D$ separately, but not jointly. It can be solved
by alternating full optimization with respect to $W$ and $D$. Each resulting sub-problem is
a very high dimensional linear program with many constraints~\citep{sandler09}, which is intractable
with standard LP solvers even for short sound signals.
In addition, convergence proofs of alternate minimization methods for NMF
typically assume strictly convex sub-problems (see \textit{e.g.} ~\citealp{tropp2003alternating}; 
\citealp{bertsekas1999nonlinear} Prop. 2.7.1), which is not the case when using non-regularized
$\ot$ as a loss.

To address this issue, \cite{rolet2016} proposed to use $\ot_\gamma$ instead,
and showed how to solve each sub-problem in the dual using fast gradient computations.
Formally, they tackle problems of the form:
\begin{equation}
\label{eq:nmf_ot}
\min_{\substack{D\in\Sigma_n^k\\
W\in\Sigma_k^t
}} 
\sum_{i=1}^t\ot_{\gamma} (\col{x}{i}, D\col{w}{i}) + R_1(\col{W}{i}) + \sum_{i=1}^k 
R_2(\col{D}{i})
\end{equation}
where $R_1$ and $R_2$ are convex regularizers that enforce non-negativity 
constraints, and $\Sigma_n$ is the $(n-1)$-dimensional simplex.

It was shown that each sub-problem of \eqref{eq:nmf_ot}
with either $D$ or $W$ fixed has a smooth Fenchel-Rockafellar dual,
which can be solved efficiently,
leading to a fast overall algorithm. However, their definition of optimal
transport
requires inputs and reconstructions to have a $\ell$-$1$ 
norm equal to $1$. This is achieved by normalizing the input beforehand, restricting 
the columns of $D$ and $W$ to the simplex, and using as regularizers
negative entropies defined on the simplex: 
\begin{equation}
    R_1(W) \coloneqq R(\rho_1, W) 
    \quad \text{and} \quad
    R_2(W) \coloneqq R(\rho_2, W)
\end{equation}
where
\begin{equation}
R(\rho, W) \coloneqq
\begin{cases}
-\rho E(W) &\mbox{ if } \|\col{W}{i}\|_1 =1,\, \forall i\\
\infty&\mbox{ otherwise.}
\end{cases}
.
\end{equation}
They showed that 
the coefficients and dictionary can be updated according to the following duality results.

\newsubsubsection{Coefficients Update}
For $D$ fixed, the optimizer of $$\displaystyle\min_{
\substack{W \in \Sigma_k^t\\
\forall i,\,D\col{w}{i} = \col{x}{i}}} \displaystyle\sum_{i=1}^t \ot_{\gamma}(\col{x}{i},D\col{w}{i}) + 
R_1(\col{w}{i})$$
~ is
\begin{equation}
\label{eq:primal_dual_W}
W^* =\left (\frac{e^{-D^\top \col{g}{i}^*/\rho_1}}{\dotp{e^{-D^\top \col{g}{i}^*/\rho_1}}{\ones}}\right )_{i=1}^m
\end{equation}
with 
\begin{equation}
\label{eq:dual_W}
\col{g}{i}^*\in\argmin{\vector{g}\in\mathbb{R}^s}\,  \ot_{\gamma}^\star (\col{x}{i}, 
\vector{g}) + R_1^\star(-D^\top \vector{g}).
\end{equation}

We can solve Problem~\eqref{eq:dual_W} with accelerated gradient 
descent~\citep{nesterov1983method}, and recover the optimal weight matrix
with the primal-dual relationship~\eqref{eq:primal_dual_W}. The
value and gradient of the convex conjugate of $R$ with respect to its second variable are:
\begin{align*}
&R^\star(\rho, \vector{x}) = \rho\log \dotp{e^{\vector{x}/\rho}}{\mathbf{1}}\\
&\nabla_x R^\star(\rho ,\vector{x}) = 
\frac{e^{\vector{x}/\rho}}{\dotp{e^{\vector{x}/\rho}}{\mathbf{1}}}.
\end{align*}

\newsubsubsection{Dictionary Update}
For $W$ fixed, the optimizer of $$\displaystyle\min_{
\substack{D \in \Sigma_m^k\\
\forall i,\,D\col{w}{i} = \col{x}{i}}}\displaystyle\sum_{i=1}^t \ot_{\gamma}(\col{x}{i}, D\col{w}{i}) + \sum_{i=1}^k 
R_2(\col{D}{i})$$
is
\begin{equation}
\label{eq:primal_dual_D}
D^* =\left(\frac{e^{-G^*\row{w}{i}^\top/\rho_2}}{\dotp{e^{-G^*\row{w}{i}^\top/\rho_2}}{\mathbf{1}}}
\right)_{i=1}^k
\end{equation}
with 
\begin{equation}
\label{eq:dual_D}
G^* \in \argmin{G \in \mathbb{R}^{n\times t}} \displaystyle \sum_{i=1}^t 
\ot_{\gamma}^\star (\col{x}{i}, \col{g}{i}) + 
\displaystyle\sum_{i=1}^kR_2^\star(-G\row{w}{i}^\top).
\end{equation}
Likewise, we can solve Problem~\eqref{eq:dual_D} with accelerated gradient 
descent, and recover the optimal dictionary matrix
with the primal-dual relationship~\eqref{eq:primal_dual_D}.

These duality results allow us to go from a constrained primal problem for which each
evaluation of the objective and its gradient requires solving an optimal
transport problem, to a non-constrained dual problem whose objective and gradient
can be evaluated in closed form. The primal constraints $\|\col{x}{i}\|_1 = \|D\col{W}{i}\|_1$
and $D\col{W}{i} \geq 0$ $\forall i$ are enforced by the primal-dual relationship.
Moreover, the use of an entropy regularization, with
$\gamma >0$, makes $\ot_\gamma$ smooth with respect to its second variable.

\section{Method}

We now present our approach for optimal transport BSS. First we introduce the changes to
\cite{rolet2016} that are necessary for computing optimal transport NMF on
STFT spectrograms of sound data. We then define a transportation cost between frequencies.
Finally we show how to reconstruct sound signals from the separated spectrograms.

\subsection{Signal Separation With NMF}

We use a supervised BSS setting similar to the one described in \cite{schmidt2006single}. For each source $k$ we have access
to training data $X^{(k)}$, on which we learn a dictionary $D^{(k)}$ with NMF
\begin{equation*}
\min_{W, D^{(k)}} \sum_{i=1}^t\ell(\col{x}{i}, D^{(k)}\col{w}{i}) + R_1(W) + R_2(D^{(k)}).
\end{equation*}

Then, given the STFT
spectrum of a mixture of voices $X$, we reconstruct separated spectrograms
$X^{(k)} = D^{(k)}W^{(k)}$ for $k=1,\dots N$ where $W^{(k)}$sare the solutions of
\begin{equation*}
\min_{W^{(1)}, \dots, W^{(N)}} \sum_{i=1}^t\ell(\col{x}{i}, \sum_{k=1}^N
D^{(k)}\col{w}{i}^{(k)}) + \sum_{k=1}^N R_1(W^{(k)}).
\end{equation*}

The separated signals are then reconstructed from each $X^{(k)}$ with the process
described in Section~\ref{sec:post-processing}.

In practice at test time, the dictionaries are concatenated in a single matrix
$D=(D^{(k)})_{k=1}^N$, and a single matrix of coefficients $W$ is learned, which we
decompose as $W=(W^{(k)})_{k=1}^N$. This allows us to focus on problems of the form
\begin{equation*}
\min_{W, D} \sum_{i=1}^t\ell(\col{x}{i}, D\col{w}{i}) + R_1(W) + R_2(D).
\end{equation*}

\subsection{Non-normalized Optimal Transport NMF}

Normalizing the columns of the input $X$, as in \cite{rolet2016},
is not a good option in the context of signal processing, since
frames with low amplitudes are typically noise and it would amplify 
them.

However, our definition of optimal transport does not require inputs
to be in the simplex, only to have the same $\ell$-$1$ norm.
With this definition,
the convex conjugate $\ot^\star$ of $\ot$ 
and its gradient still have the same value as in~\cite{cuturi2016smoothed},
and we can simply relax the condition on $W$ to be $W\geq 0$ in Problem~\eqref{eq:nmf_ot}.
We keep a simplex constraint on the columns of the dictionary $D$ so that
each update is guaranteed to stay in a compact set.
We use $R_1=-\rho_1 E$, a negative entropy
defined on the non-negative orthant as the coefficient matrix regularizer
and for $R_2$ we keep the non-negative entropy 
defined on the simplex. The problem then becomes
\begin{equation}
\label{eq:nmf_ot_relaxed}
\min_{\substack{D\in\Sigma_n^k\\
W\in\mathbb{R}_+^{k\times t}}} 
\sum_{i=1}^t\ot_{\gamma} (\col{x}{i}, D\col{w}{i}) + R_1(\col{W}{i}) + \sum_{i=1}^k 
R_2(\col{D}{i})
\end{equation}

The dictionary update is the same as in \cite{rolet2016}. However, the
coefficient updates need to be modified as follows.

\newsubsubsection{Coefficients Update}
    For $D$ fixed, the optimizer of $$\displaystyle\min_{\substack{
W \in \mathbb{R}_+^{k 
\times t}\\
\forall i,\,D\col{w}{i} = \col{x}{i}
}}\displaystyle\sum_{i=1}^t \ot_{\gamma}(\col{x}{i},D\col{w}{i}) + 
R_1(\col{w}{i})$$ 
is $W^* =\left (e^{-D^\top \col{g}{i}^*/\rho_1 - 1}\right )_{i=1}^m$, with 
\begin{equation}
\label{eq:dual_W_relaxed}
\col{g}{i}^*\in\argmin{\vector{g}\in\mathbb{R}^s}\,  \ot_{\gamma}^\star (\col{x}{i}, 
\vector{g}) + R_1^\star(-D^\top \vector{g}).
\end{equation}
The concave conjugate of $E$ and its gradient can be evaluated with:
\begin{align*}
&R_1^\star(\vector{x}) = \rho_1\dotp{e^{\vector{x}/\rho_1 - 1}}{\mathbf{1}}\\
&\nabla 
R_1^\star(\vector{x})  =e^{\vector{x}/\rho_1 - 1}.
\end{align*}

\subsection{Cost Matrix Design}
\label{sec:cost_matrix}

\begin{figure*}[h!]
\includegraphics[width=\textwidth]{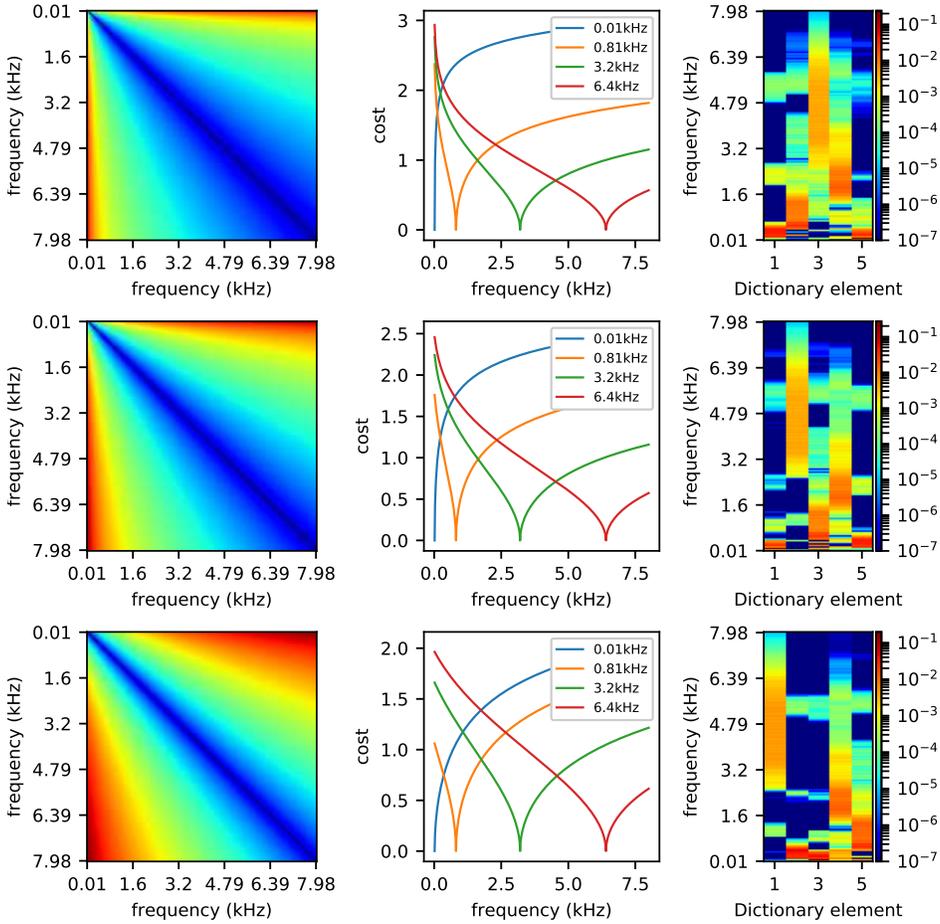}%
  \caption{\textbf{$\lambda$ parameter of the Cost Matrix.}
      Influence of parameter $\lambda$ of the cost matrix. Left: cost matrix; 
center: sample lines of the cost matrix; right: dictionary learned on the 
validation data. Top: $\lambda=1$; center: $\lambda=100$; bottom: $\lambda=1000$.}
  \label{fig:ground_metric_lambda}
\end{figure*}

In order to compute optimal transport on spectrogams and perform NMF, we need a 
cost
matrix $C$, which represents the cost of moving weight from frequencies in the 
original
spectrogram to frequencies in the reconstructed spectrogram. 
\cite{schmidt2006single} 
use the mel scale to quantize spectrograms, relying on the fact that the perceptual 
difference
between frequencies is smaller for the high frequency than for the low frequency domain. 
Following the same intuition,
we propose to map frequencies to a log-domain and apply a cost function in that
domain. 
Let $f_j$ be the frequency of the $j$-th bin in
an input data spectrogram, where $1 \le j \le m$. Let $\hat{f}_{\hat{j}}$ be the frequency of the $\hat{j}$-th bin in
a reconstruction spectrogram, where $1 \le \hat{j} \le n$. We define the cost matrix
$C \in \mathbb{R}^{m \times n}$ as
\begin{equation}
c_{j\hat{j}} = \left| \log(\lambda + f_j) - \log(\lambda + \hat{f}_{\hat{j}}) \right|^p
\end{equation}
with parameters $\lambda \geq 0$ and $p >0$.
Since the mel scale is a log scale, it is included in this definition for some parameter
$\lambda$.
Some illustrations of our cost matrix for different values of $\lambda$ are
shown in Figure~\ref{fig:ground_metric_lambda}, with $p=0.5$. It shows that 
with our definition, 
moving weights locally is less costly for high frequencies than low ones, and 
that this
effect can be tuned by selecting $\lambda$.

\begin{figure*}[h!]
\includegraphics[width=\textwidth]{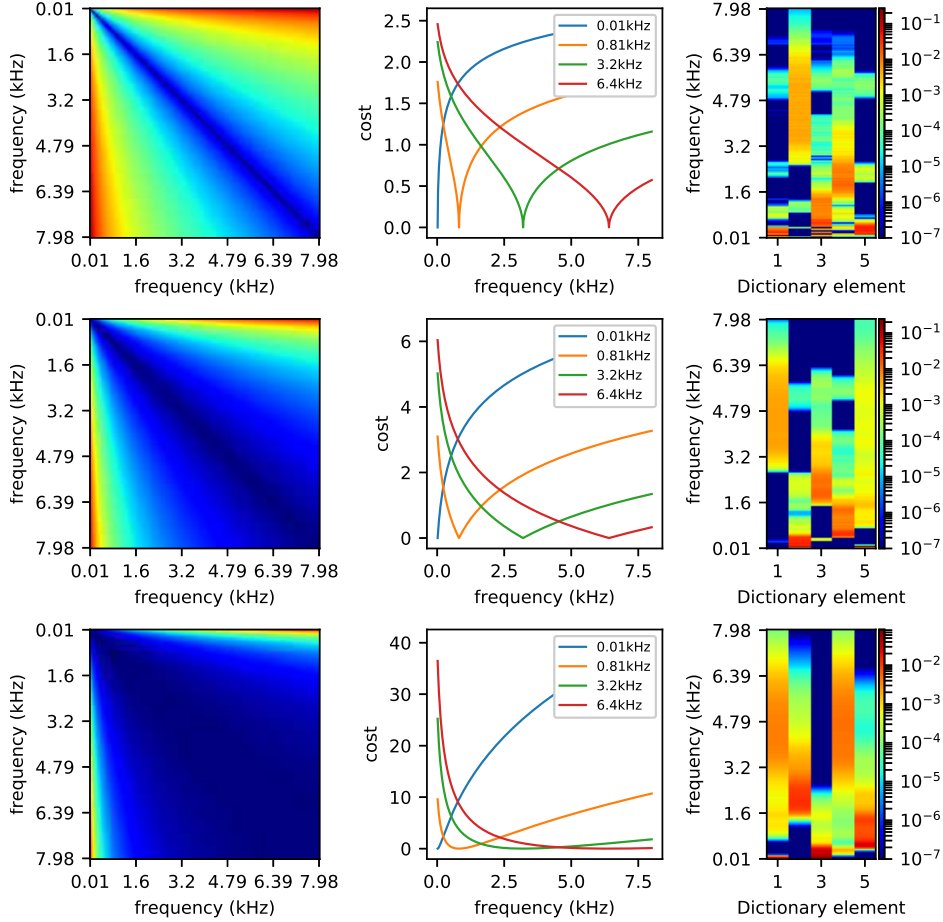}%
  \caption{\textbf{Power of the Cost Matrix.}
      Influence of the power $p$ of the cost matrix. Left: cost matrix; center: 
sample lines of the cost matrix; right: dictionary learned on the validation 
data. Top: $p=0.5$; center: $p=1$; bottom: $p=2$.}
  \label{fig:ground_metric_p}
\end{figure*}

Figure~\ref{fig:ground_metric_p} shows the effect of $p$ on the learned 
dictionaries. Using $p=0.5$ yields a cost that is more spiked, leading to dictionary 
elements that can have several spikes in the same frequency bands, whereas 
$p\geq 1$ tends to produce smoother dictionary elements.

Note that with this definition and $p\geq 1$ , $C$ is a distance matrix to the 
power $p$ when the source and target frequencies are the same. If $p=0.5$, $C$ 
is the point-wise square-root of a distance matrix and as such is a distance 
matrix itself. $\ot(.,.)^{1/p}$.

Parameters $p=0.5$ and $\lambda=100$ yielded better results for Blind Source 
Separation on the validation set and were accordingly used in all our experiments.

\subsection{Post-processing}
\label{sec:post-processing}

\newsubsubsection{Wiener Filter} In the case where the reconstruction is in the same 
frequency domain as the 
original signal, the classical way to recover each voice in the time domain is 
to apply a Wiener filter. Let $X$ be the original Fourier spectrum, $X^{(1)}$ and 
$X^{(2)}$ the separated spectra such that $X\approx X^{(1)} + X^{(2)}$. The Wiener 
filter builds $\hat{X^{(1)}} = X\odot\frac{X^{(1)}}{X^{(1)} + X^{(2)}}$ and 
$\hat{X^{(2)}} =  X\odot\frac{X^{(2)}}{X^{(1)} 
+ X^{(2)}}$, before applying the 
original spectra's phase and performing the inverse STFT.

\newsubsubsection{Generalized Filter} We propose to extend this filtering to the
case where $X^{(1)}$ and $X^{(2)}$ are not in the same domain as $X$.
This may happen for example if the test data is recorded using a different
sample frequency, or if the STFT is performed with a different time-window
than the train data. In such a case, $D^{(1)}$ and $D^{(2)}$ are in the
domain of the train data, and to are $X^{(1)}$ and $X^{(2)}$, but $X$
is in a different domain, and its coefficients correspond to different sound
frequencies. As such, we cannot use Wiener filtering.

Instead we propose to use the optimal transportation matrices to produce separated
signals $\hat{X^{(1)}}$ and $\hat{X^{(2)}}$ in the same domain as $X$.
Let $T_{(i)} \in\displaystyle
\argmin{\Pi\in U(\col{x}{i}, \col{x}{i}^{(1)}+ 
\col{x}{i}^{(2)})} 
\dotp{C}{\Pi}$. With Weiner filtering, $x_i$ is decomposed into its components
generated by $x_1^{(1)}$ and $x_2^{(2)}$. We use the same idea and separate the transport matrix $T_{(i)}$ into:

\begin{align*}
T^{(1)}_{(i)}&= T_{(i)} \diag{\frac{\col{x}{i}^{(1)}}{\col{x}{i}^{(1)}+ 
\col{x}{i}^{(2)}}}\\
T^{(2)}_{(i)}&= T_{(i)} \diag{\frac{\col{x}{i}^{(2)}}{\col{x}{i}^{(1)}+ 
\col{x}{i}^{(2)}}}
\end{align*}

$T^{(1)}_{(i)}$ (resp. $T^{(1)}_{(i)}$) is a transport matrix between $\frac{\col{x}{i}^{(1)}}{\col{x}{i}^{(1)}+ 
\col{x}{i}^{(2)}}$ (resp. $\frac{\col{x}{i}^{(2)}}{\col{x}{i}^{(1)}+ 
\col{x}{i}^{(2)}}$) and $\hat{\col{x}{i}}^{(1)}$ (resp. $\hat{\col{x}{i}}^{(2)}$), where

\begin{align*}
\hat{\col{x}{i}}^{(1)} &= T^{(i)} \frac{\col{x}{i}^{(1)}}{\col{x}{i}^{(1)}+ \col{x}{i}^{(2)}}\\ 
\hat{\col{x}{i}}^{(2)} &= T^{(i)} \frac{\col{x}{i}^{(2)}}{\col{x}{i}^{(1)}+ \col{x}{i}^{(2)}}
\end{align*}

Similarly to the classical Wiener filter, we have
\begin{align*}
\hat{\col{x}{i}}^{(1)}  + \hat{\col{x}{i}}^{(2)} &= T^{(i)} 
\frac{\col{x}{i}^{(1)}}{\col{x}{i}^{(1)}+ \col{x}{i}^{(2)}} + T^{(i)} 
\frac{\col{x}{i}^{(2)}}{\col{x}{i}^{(1)}+ \col{x}{i}^{(2)}}\\ 
&=T^{(i)}\ones\\
&=\col{x}{i}
\end{align*}

\newsubsubsection{Heuristic Mapping} As an alternative to this generalized filter,
we propose to simply map the reconstructed signal to the same domain as $X$ by assigning
the weight of a $\hat{f}_j$ in a spectrogram to its closest neighbor in $(f_i)_{i=1}^n$,
according to the distance we defined for the cost matrix (see Section~\ref{sec:cost_matrix}).

\newsubsubsection{Separated Signal Reconstruction} Separated sounds are reconstructed by inverse
STFT after applying a Wiener filter or generalized filter to $X^{(1)}$ and $X^{(2)}$.

\section{Results}

In this section we present the main empirical findings of this paper. We start by describing the dataset that we used
and the pre-processing we applied to it. We then show that the optimal transport
loss allows us to have perceptually good reconstructions of single voices, even with few dictionary elements.
Finally we show that the optimal transport loss improves upon a Euclidean loss for BSS with an NMF
model, both in single-domain and cross-domain settings.

\subsection{Dataset and Pre-processing}

We evaluate our method on the English part of the Multi-Lingual Speech Database
for Telephonometry 
1994 dataset\footnote{\dataseturl}. The data 
consists of 
recordings of the voice of four males and four females pronouncing each 24 
different English sentences. We split each person's audio file time-wise into 
$25\%$-$75\%$ train-test data. The files are re-sampled to $16kHz$ and treated 
as mono signal.

One of the male voices and one of the female voices are only used for hyper-parameter selection,
and are not included in the results.

The signals are analysed by STFT with a Hann window, and a window-size of $1024$, leading to $513$
frequency bins ranging from $0$ to $8$kHz. The constant coefficient is removed from 
the NMF analysis and added for reconstruction in post-processing.

Hyper-parameters are selected on validation data consisting if the first male 
and female voice, which are excluded from the evaluation set.

Initialization is performed by setting each dictionary column to the optimal
transport barycenter of all the time frames of the training data, to which we added
Gaussian noise (separately for each column). The barycenters are computed using the
algorithm of \cite{benamou2015iterative}.

\subsection{NMF Audio Quality}

We first show that using an optimal transport loss for NMF leads to better 
perceptual reconstruction of voice data. To that end, we evaluated the PEMO-Q 
score~\citep{huber2006pemo} of isolated test voices. The dictionaries are 
learned on the isolated voices in the train dataset, and are the same as in the 
following separation experiment.

\begin{figure*}[h!]
\includegraphics[width=\textwidth]{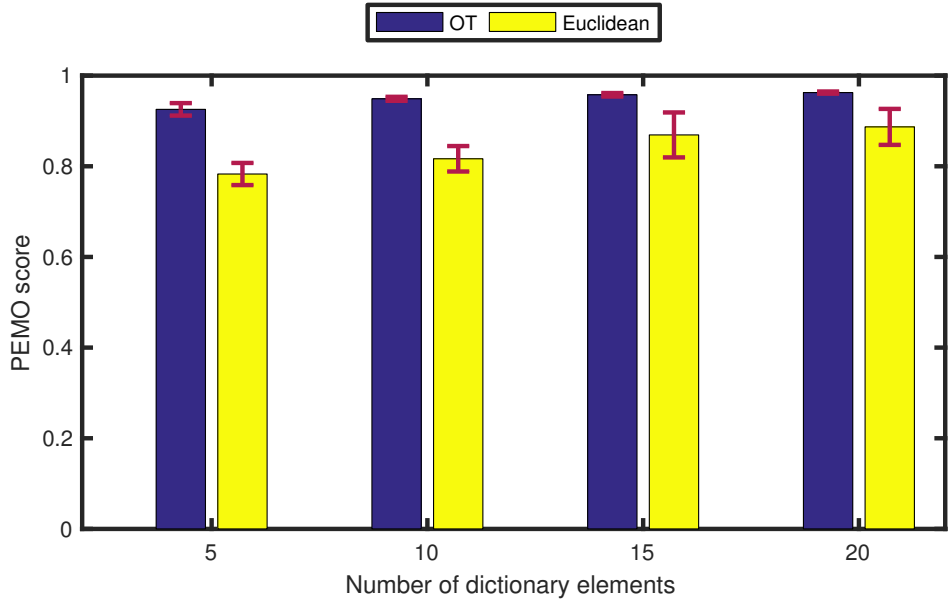}
  \caption{\textbf{Perceptive Quality Score.}
      Average and standard deviation of PEMO scores of non-mixed voices with 
optimal transport (blue) or Euclidean (yellow) NMF.}
\label{fig:NMF_quality}
\end{figure*}

Figure~\ref{fig:NMF_quality} shows the mean and standard deviation of the scores for 
$k\in\{5,\,10,\,15,\,20\}$ with optimal transport and Euclidean NMF. The 
PEMO-Q score of optimal transport NMF is significantly higher for any value 
of $k$. We found empirically that other scores such as SDR or SNR tend to be better for the Euclidean NMF, even
though the reconstructed voices are clearly worse when listening to them (see
additional files 1 and 2). Optimal transport can reconstruct clear and
intelligible voices with as few as $5$ dictionary elements.

\subsection{Blind Source Separation}

We evaluate our Blind Source Separation using the PEASS score proposed in 
\cite{emiya2011subjective}, which they claim is closer to how humans would score
BSS than SDR. We only consider mixtures of two voices, where the mixture is simply
an addition of the sound signals.

\newsubsubsection{Single-Domain Blind Source Separation}%
We first show that using an optimal transport NMF improves on Euclidean NMF for 
BSS
using the same frequencies in the spectrogram of the train and test data.
In this experiment, both the training and test data are processed in exactly the same
way, so that at train and test time $(f_i)_i = (\hat{f}_i)_i$. For 
Euclidean-based BSS, we reconstruct the signal using a Wiener filter before 
applying inverse
STFT. For optimal transport-based source separation, we evaluate separation using
either the Wiener filter or our generalized filter.

\begin{figure*}[h!]
\includegraphics[width=\textwidth]{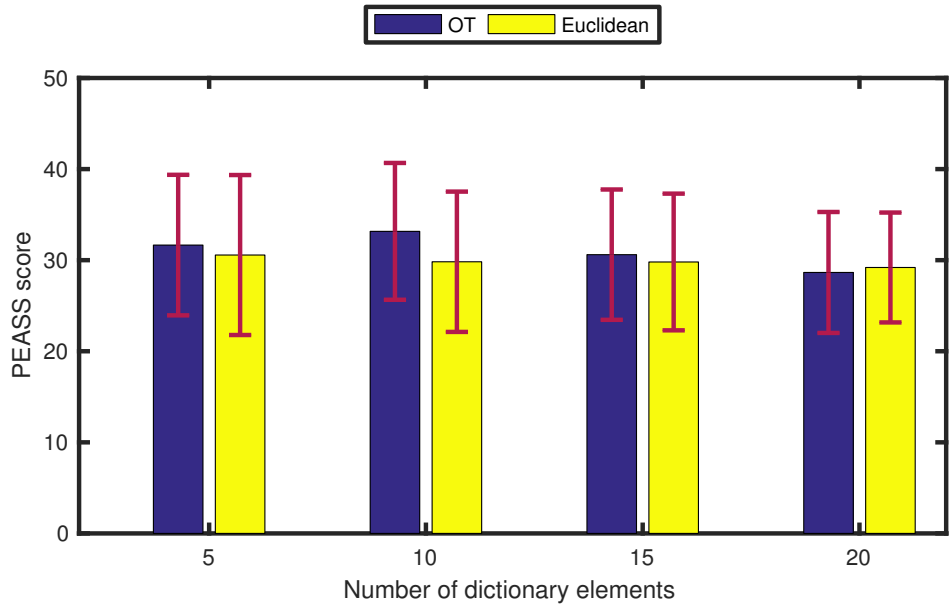}
  \caption{\textbf{Average Separation Score.}
      Average and standard deviation of PEASS scores with optimal transport 
(blue) or 
Euclidean (yellow) NMF, both reconstructed with the Wiener filter.}
\label{fig:separation_score}
\end{figure*}

Figure~\ref{fig:separation_score} shows mean and standard deviation of the PEASS scores
for $k\in\{5,\,10,\,15,\,20\}$. The scores
are higher with $k=5$ or $k=10$ and in both cases optimal transport yields 
better results.

\begin{figure*}[h!]
\includegraphics[width=\textwidth]{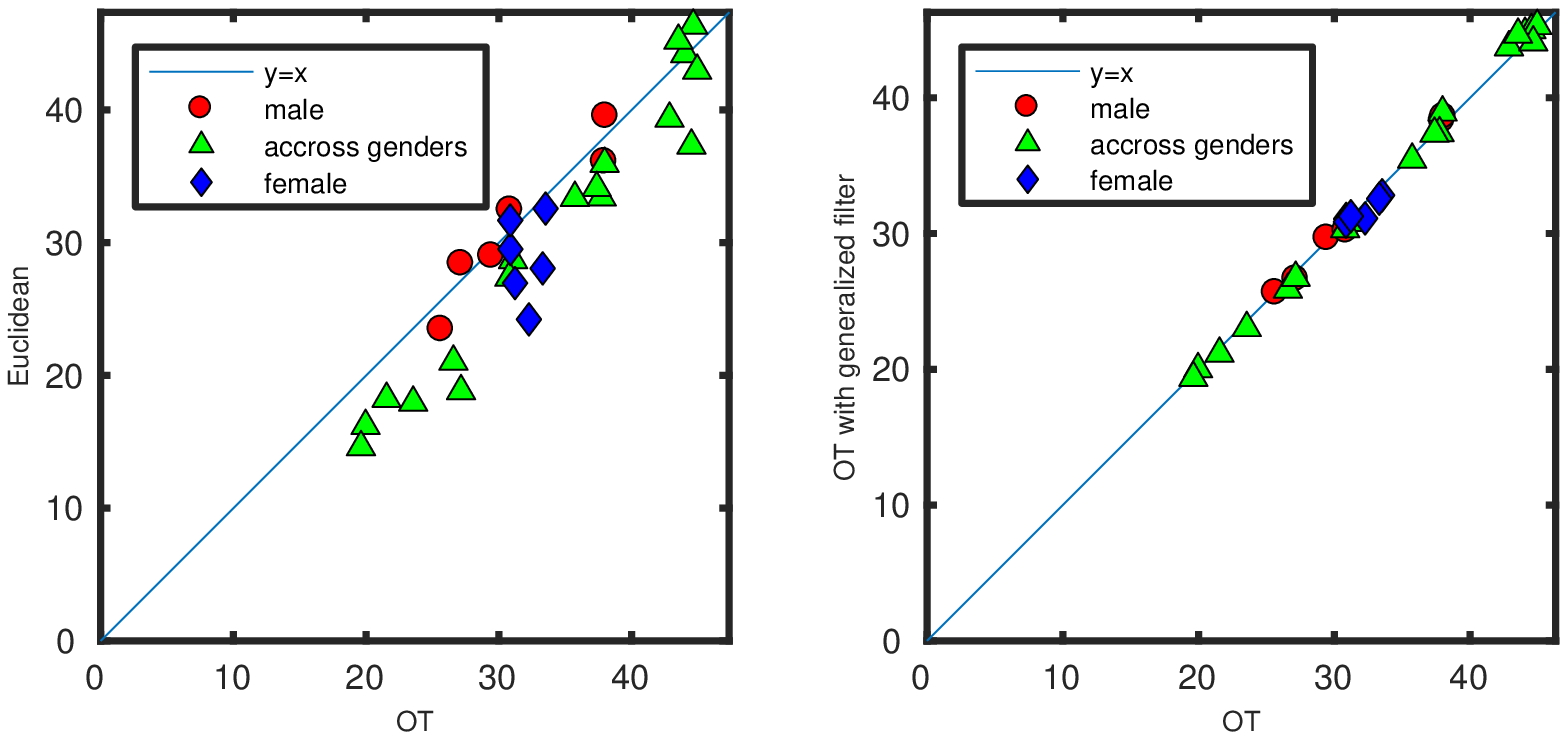}%
  \caption{\textbf{Single Domain Separation Score.} Comparison between optimal transport NMF and Euclidean NMF(left) or 
  optimal transport NMF
with generalized
  Wiener filter (right). Each data-point represents the PEASS scores of one file 
when mixed with another,
  where the x coordinate is the optimal transport with Wiener filter's score and 
the y coordinate is the
  score of the compared method.}
\label{fig:separation_score_scatter}
\end{figure*}

Figure~\ref{fig:separation_score_scatter} shows a comparison for each pair of 
mixed voices, with $k$
selected on the validation set ($k=5$ for Euclidean and $k=10$ for optimal transport NMF). It
shows that the PEASS score is better with an optimal transport loss for almost 
all files. We
can further see that in the case of single domain BSS, the Wiener filter and our 
generalized Wiener
filter yields very similar results.

\newsubsubsection{Cross-Domain Blind Source Separation} In this experiment, we
keep the dictionaries trained for the single domain experiment, but we re-process
the test data with a different time-window of $600$ for the STFT.
Although $(f_i)_i \neq (\hat{f}_i)_i$, we can still compute optimal transport between the
spectrograms thanks to our cost matrix.

\begin{figure*}[h!]
\includegraphics[width=\textwidth]{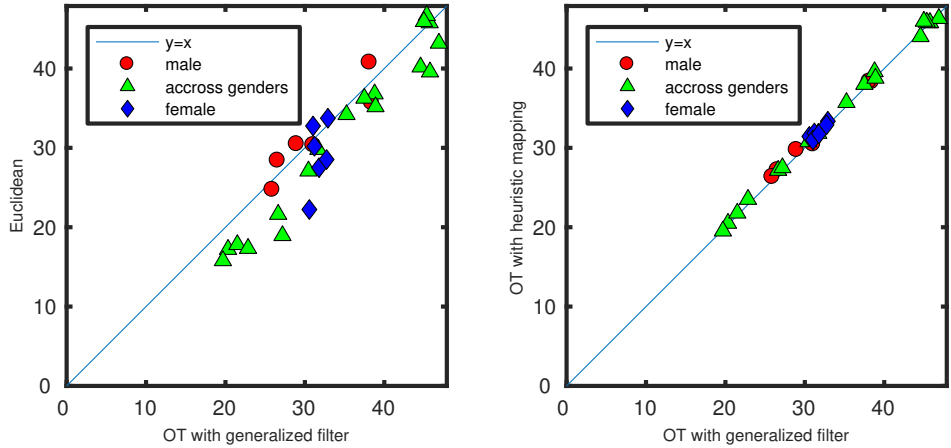}%
  \caption{\textbf{Cross Domain Separation Score.} Comparison between optimal transport NMF with generalized
  Wiener filter and Euclidean NMF (left) or
  optimal transport NMF with heuristic mapping (right) on the cross domain speech separation task.
   Each data-point represents the PEASS scores of one file 
when mixed with another,
  where the x coordinate is the optimal transport with generalized Wiener filter's score and 
the y coordinate is the
  score of the compared method.}
\label{fig:cross_domain_separation_score_scatter}
\end{figure*}

Figure~\ref{fig:cross_domain_separation_score_scatter} shows the resuts on the train
set. The score for Euclidean NMF is computed by first mapping the test data to the
same domain as the train data, using heuristic mapping, and then performing same-domain
separation. Both the heuristice mapping and generalized filter improve upon using Euclidean
NMF, and they both achieve similar results. Still, the use of our generalized filter
allows to have the exact same processing whether performing single domain or cross
domain separation, the only difference being the cost matrix $C$,
while the heuristic mapping requires additional post-processing
and also requires to choose rules for the mapping.

\section{Discussion}

\newsubsubsection{Regularization of the Transport Plan}
In this work we considered entropy-regularized optimal transport as introduced by
\citet{cuturi13}. This allows us to get an easy-to-solve dual problem since its
convex conjugate is smooth and can be computed in closed form. However, any convex
regularizer would yield the same duality results, and could be considered as long as its
conjugate is computable. For instance, the squared $L^2$ norm regularization was considered
in several recent works \citep{blondel2017smooth, seguy2017large} and was shown
to have desirable
properties such as better numerical stability or sparsity of the optimal transport plan. Moreover, 
similarly to entropic regularization, it was shown that the convex conjugate and
its gradient can be computed in closed form \citep{blondel2017smooth}.

\newsubsubsection{Learning Procedure}
Following the work of \cite{rolet2016}, we solved the NMF problem with an
alternating minimization approach,
in which at each iteration a complete optimization is performed on either the dictionary
or the coefficients. While this seems to work well in our experiments, it would be
interesting to compare with smaller steps approach like in \cite{lee2001algorithms}.
Unfortunately such updates do not exist to our knowledge: gradient methods in
the primal would be prohibitively slow, since they involve solving $t$ large optimal transport
problems at each iteration.

\section{Conclusion}

We showed that using an optimal transport based loss can improve performance of NMF-based
models for voice reconstruction and separation tasks. We believe this is a first step towards using optimal
transport as a loss for speech processing, possibly using more complicated models such 
neural networks.
The versatility of optimal transport, which can compare spectrograms on different frequency domains,
lets us use dictionaries on sounds that are not recorded or processed in the same way as the training set.
This property could also be beneficial to learn common representations (\textit{e.g.} dictionaries) for different
datasets.

\section*{Additional Files}

All of the additional files are wav files.
  
  \subsection*{Additional file 1 --- Reconstruction with optimal transport NMF}
  
This file contains the reconstructed signal for 6 test sentences
of the male validation voice with optimal transport NMF and
a dictionary of rank 5 (5 columns), where the dictionary was learnt on the training sentences of
the same voice.

  \subsection*{Additional file 2 --- Reconstruction with Euclidean NMF}

This file contains the reconstructed signal for 6 test sentences
of the male validation voice with Euclidean NMF and
a dictionary of rank 5 (5 columns), where the dictionary was learnt on the training sentences of
the same voice.

\section*{Acknowledgements}

The authors would like to thank Arnaud Dessein, who gave helpful insight on the cost matrix design.

\bibliographystyle{plainnat} 
\bibliography{paper}     
\end{document}